\documentclass{sig-alternate}
\usepackage{times,amsmath,amssymb,epsfig}
\usepackage{graphicx}

\usepackage{subfigure}

\usepackage{url}
\urldef{\mailsa}\path|{alfred.hofmann, ursula.barth, ingrid.haas, frank.holzwarth,|
\urldef{\mailsb}\path|anna.kramer, leonie.kunz, christine.reiss, nicole.sator,|
\urldef{\mailsc}\path|erika.siebert-cole, peter.strasser, lncs}@springer.com|    

\newcommand{\x}{\ensuremath{x}}

\title{Principle of Need-to-Act}

\numberofauthors{1}
\author{%
\alignauthor
Ashish Kundu\\ \affaddr{IBM T J Watson Research Center}\\
\affaddr{Yorktown Heights, New York, USA}\\
      \email{akundu@us.ibm.com}
}
\begin{document}
\maketitle
\begin{abstract}
In this paper, we have introduced the notion of ``Principle of Need-to-Act''.
This principle is essential towards developing secure systems, security
solutions and analyzing security of a solution. 
 
\end{abstract}
%


\section{Introduction}
In computer security including cyber-physical security,  a number of principles have been in introduced and have been applied in practice~\cite{infosecprinciples}. While they cover most practices that are implemented today, they do not cover certain weaknesses that a system can be vulnerable to.

A system is designed to receive a request/input and then process that request, which is a form of action in terms of reaction to the input. An action is any task that is carried out by the system. 

If a system can be made to "act" by an entity X within or outside the system, even though the system does not need to act in order to provide a functionality, then that is a weakness, and the system is vulnerable to attacks by or via entity X. 

If X is within the system, X is an insider threat, and likewise if X is outside the system, X is an external threat. The type of attack can be a DoS attack, or an energy attack, or simply part of a multi-stage attack -- exploiting "lack of enforcement of need-to-act principle" may overwhelm the system capabilities and change the security behaviour of the system allowing the attacker to exploit another vulnerability as the next/later step. 

SYN-flooding vulnerability arises from the fact that the network system (subsystem of kernel) did not implement "need-to-act" -- there was no need to act on such requests by the system. The act was to  (1) open a socket and (2) wait for the SYN packet to arrive. 

The question arises whether need-to-act can be effectively implemented. In general, an informal argument is -- it can be easily argued that it is undecidable to prove any program implements need-to-act principle.

\section{Principle of Need-to-Act}
In order to prevent energy-attacks~\cite{Wu:2011:EAS:2028052.2028060}, it is
essential that a computing entity enforces the principle of need to act. The
idea is based on the fact that energy is consumed everytime an action is carried out. 
An action can be a memory access, or computation, or network transfer or any other action carried out by a computing
device. If energy attacks are to be prevented, then only the actions that are
"essential" to the desired functionality are to be carried out and no other
actions must be carried out. \\

The principle of Need to Act is different than the
principle of ``need to know'' or other basic security
principles~\cite{infosecprinciples}. In the need to know principle, an entity
gets access to know some information if and only if, the entity has a legitimate
"need" to know that information. Need to act principle states the following:\\
\noindent
"An  action is carried out by an entity if and only if the action needs to be
carried out by that entity in order to compute/deliver the desired functionality or
output".\\

Need-to-know protects knowledge or information, whereas, ``need-to-act" protects
computing resources as well as knowledge, which accessed by some ``act''. As
with need-to-know, the notion of "need" is defined by the context of
computation, and the time. In the context of need-to-know, an action is not
needed anymore if the context is not valid anymore. Need-to-know when evaluated
requires certain action to be carried out. Therefore, it does not cover the
``need to act'' principle.
The question is how do we enforce "need-to-action". For  all
security-related operations including "need-to-know", we often employ the
"principle of mediation": there is an entity that decides whether the requestor to some information has
legitimate access to that specific information in that context of computing.
This intermediary evaluates some conditions preceding allowing or disallowing
access. Can we apply such an intermediary or a set of preconditions to enforce
"need-to-act"?

There is indeed a notion of a precondition that needs to be satisfied before
carrying out an action. However, checking whether a precondition is satisfied by
the context of the action requires some other actions to be carried out such as
processing the boolean conditions stated as part of the precondition. This
checking itself is a set of actions. If an attacker succeeds even in getting the
computer process the boolean conditions, it has been successful in mounting an
energy attack; the impact of the attack may be miniscule with respect to energy
consumption.
\begin{figure*}
\centering
\includegraphics[scale=0.4]{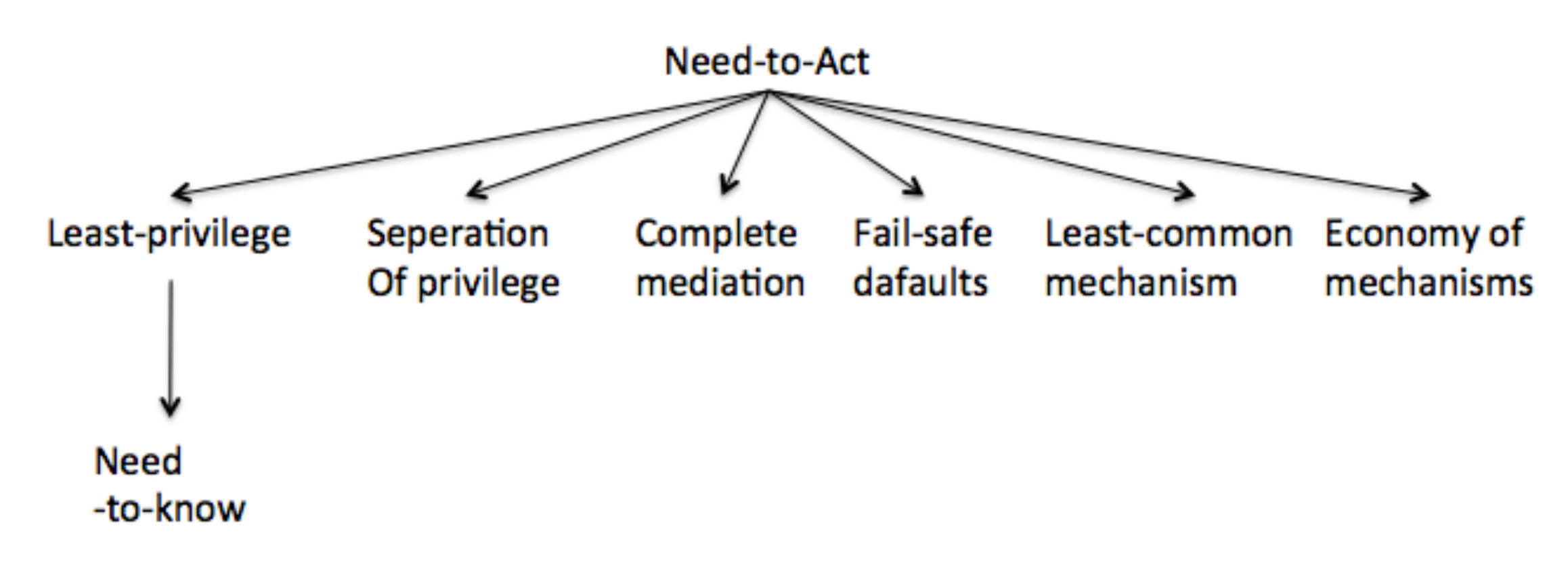}
\caption{Need-to-act and other principles of security}
\label{fig:n2a}
\end{figure*}

Therefore, employing an intermediary to check whether "need-to-act" is valid or
invalid with the objective to prevent or avoid energy attacks is itself
self-defeating -- it leads to energy consumption due to "checking the
conditions". However, we could employ two different computing entities with
different sources of energy: one entity carries out the checks, and the other
one carries out the actions that are needed to provide the functionality
desired. The first entity is vulnerable to energy attacks, and the second one is
not, if the first one is a "perfect enforcer of need-to-act" for the second.
Therefore, a practical implementation of "need-to-act" would rely on a
distributed implementation, where there are different energy sources, and the
one that enforces need-to-act is protected from energy attacks (which maybe
explained as a recursive requirement, at least that is true in theory).

Almost all programs are suboptimal, and thus would fail with need-to-act
principle. Similarly almost all processes do not also support need-to-act.
However, formulation of this principle in a conceptual manner is one goal of
this paper, and the other goal of this paper is how to implement this principle
in a practical manner. 

For reasons mentioned above, need-to-act cannot be  implemented in
its true sense, primarily because some action is needed to be taken in order to differentiate between
legitimate and illegitimate accesses, requests or computations. A practical
implementation has the following characteristics: (1) any action $A(X)$ taken by
system X in response to a request/action $A(Z)$ by Z in order to verify if
$A(Z)$ is a legitimate request (authorized) and whether any processing of $A(Z)$
by X is needed, then the cost of receiving/processing of $A(X)$ by $X$ should be
limited to certain static or dynamic values. (2) If entity X has to be
protected from attackers Z, then a separate entity Y whose value is less than X
implements the need-to-act principle. Such a strategy prevents X to be attacked
directly such as with energy attacks.

Where can this principle be implemented? This principle can be implemented at
any level of a computing system -- at the hardware level, at the software level
or at the firmware level. Any system that takes an action and if such an action
can be used against that system, then need-to-act policy should be implemented.
Obfuscation of such actions can be carried out in order not to leak any
information or not to carry out an action that can harm the system. Such
obfuscation can be adaptive based on the type of the component that is asking to
carry out such actions, or the context and state of the system. Even if the
system is centralized, distributed, disaggregated or even a miniature system
such as IoT device, any action may not be carried out as requested or the
frequency of the action maybe obfuscated.

Figure~\ref{fig:n2a} presents the relation of need-to-act with other security
principles as proposed by Saltzer and Schroeder~\cite{infosecprinciples}.
Need-to-act is the most versatile principle that covers all other principles.

What it implies is that if we implement one of the principles already defined in
the literature, we shall be implementing some form of need-to-act. The principle
of need-to-act is an over-arching principle.

\bibliographystyle{IEEEtran}
\bibliography{main}

\end{document}